\documentclass[prc,showpacs,twocolumn]{revtex4}

\usepackage{comment}
\usepackage{epsfig}
\usepackage{graphicx}
\usepackage{amsmath}
\usepackage{dcolumn}

\begin{document}

\title{Scaling and Interference in the Dissociation of Halo Nuclei}

\author{M. S. Hussein, R. Lichtenth\"aler,}
\affiliation{Instituto de Fisica, Universidade de S\~ao Paulo, C.P.
55318, 05315-970, S\~ao Paulo, SP, Brazil}
\author{F. Nunes, }
\affiliation{National Superconducting Cyclotron Laboratory,
Michigan State University, East Lansing, Michigan 48824-1321}

\author{and I.J. Thompson}
\affiliation{Physics Department, University of Surrey, Guildford,
Surrey, GU2 7XH, U.K.}

\date{\today}
\begin{abstract}

The dissociation of halo nuclei through their collision with light and heavy
targets is considered within the Continuum Discretized Coupled Channels theory.
We study the one-proton halo nucleus $^8$B and the one-neutron halo nucleus
$^{11}$Be, as well as the more normal $^7$Be. The procedure previously employed
to extract the Coulomb dissociation cross section by subtracting the nuclear
one is critically assessed, and the scaling law usually assumed for the target
mass dependence of the nuclear breakup cross section is also tested.  It is
found that the nuclear breakup cross section for these very loosely bound
nuclei does indeed behave as $a+bA^{1/3}$. However, it does not have the
geometrically inspired form of a circular ring which seems to be the case for
normal nuclei such as $^{7}$Be. We find further that we cannot ignore
Coulomb-nuclear interference effects, which may be constructive or destructive
in nature, and so the errors in previously extracted B(E1) using the
subtraction procedure are almost certainly underestimated.
\end{abstract}
\pacs{25.60.Dz, 25.70.De, 24.10.Eq}

\maketitle


The study of  electromagnetic dissociation of halo nuclei is an important area
that supplies invaluable information about their multipole responses and
consequently the details of their structure \cite{bert88,bert2001}. The theory
usually employed for the purpose is the relativistic Coulomb excitation theory
developed by Winther and Alder \cite{bert88,winther79} which hinges on the
Fermi-Weisacker-Williams (FWW) virtual photon method. Since the measurement of
the dissociation cross section supplies the combined Coulomb and nuclear
contributions, one is forced to subtract the latter. The common prescription
employed for this subtraction procedure is the so-called scaling law: the
nuclear breakup cross section should scale linearly with the radius of the
target, and thus by measuring the cross section for the breakup of the halo
nucleus in the predominantly nuclear field of a light target, one attempts to
extrapolate to heavy targets assuming the validity of the scaling law. The
Coulomb dissociation cross section for the halo nucleus on the heavy target is
then simply obtained by subtracting from the experimental cross section an
extrapolated nuclear one calculated according to some prescription. It is this
``nuclear-free" cross section which is fitted by the FWW result in order to
extract the $B(EL)$ distribution
\cite{naka99,datta03,hansen95,tani99,glas98,aumann99,sack93,shim95,zinser97,naka94,leist01,iwasa,schumann,naka04}.

Two factors convinced us to critically assess this procedure. Several
theorists \cite{dasso98,dasso99,nag01,chat02,typel01,madd01} have recently cast
doubt on the relative importance of the nuclear contribution to the
dissociation cross section, claiming in some cases that this contribution can
be significantly larger than the Coulomb contribution in as heavy a target as
lead, invalidating the scaling law. The second important factor is the need to
supply a quantitative assessment of the Coulomb-nuclear interference terms in
the cross section.

The purpose of this letter is to settle these issues by performing a full
Continuum Discretized Coupled Channels calculation for $^8$B, $^{11}$Be and
$^7$Be dissociations in the fields of light-, medium- and heavy-mass targets at
three laboratory energies where data are available.


Coupled channels calculations were therefore performed to calculate the elastic breakup
(also called diffraction dissociation) arising in a three-body model consisting
of a two-body projectile incident on an inert target \cite{fresco,aust87}.
Between the target and each of the two projectile components we specify optical
potentials whose imaginary parts describe the loss of flux to channels beyond
elastic breakup. We calculate all orders of the tidal effects of the optical
potentials as they deform and break up the projectile.

For both  $^8$B and $^{11}$Be, projectile states were included for relative
motion up to partial waves $\ell_{\rm max}=3$ and energy $\epsilon_{\rm
max}=10$ MeV. This energy range was divided into 20 bins when $\ell=0$ for
$^8$B and when $\ell=0, 1$ for $^{11}$Be. There were 10 bins when $\ell=1, 2$
for $^8$B and  $\ell=2$ for $^{11}$Be, and 5 bins in the remaining partial
waves, all evenly spaced in momentum $k$.   The coupled channels for the
scattering of the projectile on the target were solved up to $R_{\rm max} =
500$ fm, and for partial waves up to $KR_{\rm max}$ where $K$ is the wave
number for the incident beam.  The optical potentials have the same parameters
for all energies, as given in  Table \ref{tab:potls}, using radii calculated
with an $A^{1/3}$ contribution from the target mass number. We thus impose a
regular target scaling in the radial geometry of the component-target
potentials, in order to examine the contributions from varying target size and
varying dynamical conditions.
\begin{table}
\caption{
\label{tab:potls} Optical and binding potential parameters
for $^8$B and $^{11}$Be projectiles.}
\begin{tabular}{c|ccc|ccc}
\hline
Pair & $V$ &  $r_0$ &  $a$ &  $W$ & $r_i$ &  $a_i$ \\
     & MeV &    fm    &   fm  &    MeV  &  fm&  fm\\
\hline
p+T  &  46.979&  1.17     &   0.75   & 6.98  &  1.32 &0.60  \\
$^{7}$Be+T  &  114.2 &  1.0      &  0.85   & 9.44 & 1.30   & 0.81 \\
n+T  &  37.14 &  1.17     &  0.75   & 8.12  & 1.26 & 0.58\\
$^{10}$Be+T  &  46.92 &  1.204     &   0.53   & 23.5 & 1.33  & 0.53 \\
\hline
n+$^{10}$Be $s$ &  51.51 &   1.39    &   0.52   &       &       &    \\
partial waves $pdf$ & 28.38  &   1.39    &   0.52   &  $V_{so}$ & $r_{so}$ &  $a_{so}$\\
p+$^{7}$Be  &  44.675&  1.25     &   0.52   &     4.9   &1.25  & 0.52 \\
\end{tabular}
\end{table}

\begin{figure}
\includegraphics[width=0.48\textwidth]{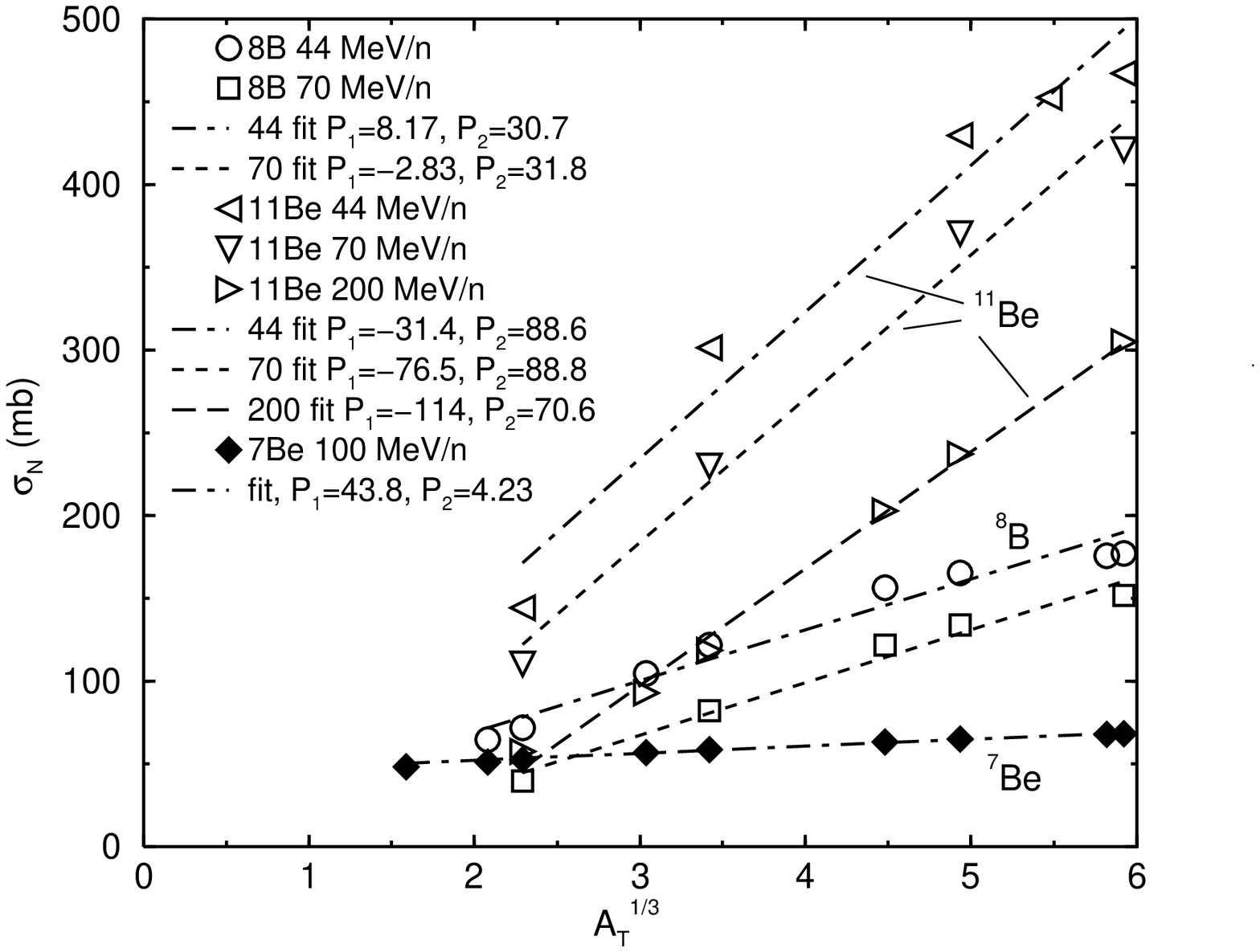}
\caption{\label{Fig1} Elastic nuclear breakup cross-section for $^8$B,
$^{11}$Be and $^{7}$Be projectiles at the indicated energies, as a function of target
mass number $A_T$, along with linear fits.}
\end{figure}

The results of the integrated nuclear breakup cross sections obtained from the
CDCC calculations are shown in figure \ref{Fig1}: circles and
squares for $^8$B at $E_{lab}=44$ and 70 MeV/n respectively,
left-triangle, down-triangle and right triangle for $^{11}$Be at
$E_{lab}=44$, 70, and 200 MeV/n respectively, and
full diamonds for $^7$Be at $100$ MeV/n (after \cite{filomena}).
These are plotted as functions
of the cubic root of the mass of the target nucleus, proportional to the
target's radius.
They are all rather accurately accounted by a linear dependence on
$A^{1/3}_T$, but with varying coefficients. We might expect that the total
direct reaction cross-section is proportional to the area of a
circular ring given \cite{acquadro80} by:
\begin{equation} \label{a+a}
\sigma_D \approx 2 \pi a(R_P+R_T)
\end{equation}
where $a$ is approximately the diffuseness of the optical potential that
describes the elastic scattering, and $R_P+R_T$ is the sum of the projectile
and target radii. In \cite{acquadro80}, it was found the Eq.~(\ref{a+a})
describes the data quite well. Guided by  Eq.~(\ref{a+a}) we anticipate the
following form of the nuclear breakup cross-section:
\begin{equation} \label{p1+p2}
\sigma_N=P_1+P_2  A_T^{1/3}
\end{equation}
where the parameters $P_1$ and $P_2$ depend on the projectile, and may also
depend on the bombarding energy and the structure of the target. We performed
linear fits to the nuclear breakup cross sections as a function of $A^{1/3}$ to
obtain the values for $P_1$ and $P_2$ shown in the legends of Figs. \ref{Fig1}.

The nuclear breakup cross section calculated with CDCC for  $^8$B and $^{11}$Be
do show the $A_T^{1/3}$ dependence of Eq.~(\ref{p1+p2}) as seen in figures
\ref{Fig1} but do not always follow the form given by
Eq.~(\ref{a+a}), in particular as, in most cases,  the cross section fits have $P_1
< 0$. At this point we should mention that our results disagree completely with
those of \cite{nag01}, where the nuclear breakup cross section was found to
scale as $A_T$. (In fact such a behaviour could only be possibly true for a very weakly
interacting system where the whole volume of the target could be effective.)

This is not the case for $^7$Be, a normal non-halo nucleus, where scaling holds
and Eq.~(\ref{a+a}) is fully satisfied,  where both $P_1$ and $P_2$ are positive
as seen in figure \ref{Fig1}. The different behaviour in the scaling of the
nuclear breakup cross-sections for $^8$B and $^{11}$Be could well be a
manifestation of their halo nature.

The fact that the nuclear cross sections do not exactly satisfy Eq.~(\ref{a+a})
with $P_1$ proportional to the radius of the projectile leads us to conclude
that nuclear breakup must be estimated from realistic dynamical models, not
from Eq.~(\ref{a+a}). As well as the present CDCC model, Glauber few-body
procedures can be used, as done in refs.~\cite{1,2,3}.

The analysis of experimental data \cite{1,2,3} started with the following
expression for the breakup cross section:
\begin{equation} \label{gsi-c+n}
\frac{d \sigma}{ d E^*} = S \frac{d \sigma_C}{ d E^*}
         + L(A_T) \frac{d \sigma (^{12}C)}{d E^*},
\end{equation}
where $S$ is the ground state spectroscopic factor, $E^*$ is the excitation
energy. The $d \sigma_C/d E^*$ is from the Coulomb FWW virtual photon formula, and the
$d \sigma (^{12}C)/d E^*$ is what is observed for a $^{12}$C target.
The $L(A_T)$ is a scaling factor, which may be determined either by fitting
Eq.~(\ref{gsi-c+n}) to the data, exploiting the different shapes of the excitation
spectra, or by calculating
\begin{equation}
L(A_T) = \sigma^{\rm th}_N(A_T) / \sigma^{\rm th}_N(12)
\end{equation}
using eikonal calculations of nuclear breakup $\sigma^{\rm th}_N(A_T)$. The
$^{12}$C is simply a reference nucleus. Of course the above incoherent sum
ignores completely  Coulomb-nuclear interference effects, which we discuss
further below.

In \cite{1,2} the reaction $^{11}$Be + $^{208}$Pb at $E = 520$ MeV/n is
considered. By adjusting $S$ and $L$, these authors obtain $L(208) = 5.6 \pm
0.4$. This is close to our value of $L$, which can be extracted from Fig.
\ref{Fig1} at $E = 200$ MeV/n, namely $L = 5.9$. Our results also agree with
the eikonal calculations of \cite{4}. In fact, at $E = 200$ MeV/n, it has been
further confirmed to us \cite{5} that within the eikonal formalism of \cite{4},
the value of $L$ comes out about 5.5, very close to our model result.

On the other hand, using the same incoherent cross-section formulae, Ref.
\cite{3} analyses the reaction $^{11}$Be + $^{208}$Pb at $E = 70$ MeV. These
authors found for $L$ the value $2.1 \pm 0.5$, much smaller than our result $L$
= 3.82 (Fig. \ref{Fig1}) but in accordance with the geometrical value obtained
from Eq.~(\ref{a+a}), which we have shown  to be not valid for halo nuclei for
at least some halo nuclei. It is conceivable that the reason for this
discrepancy resides in the neglect of the interference terms, which, as we see
next, can be rather large.
 %

\begin{figure}
\includegraphics[width=0.31\textwidth]{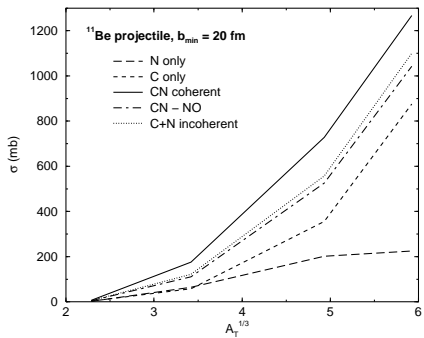}
\caption{\label{Fig4}Total  breakup, Coulomb only, and nuclear
only contributions for  $^{11}$Be projectile as a function of $A_T^{1/3}$,
for a lower radial cutoff $b_{\rm min}=20$ fm.}
\end{figure}

\begin{figure}
\includegraphics[width=0.31\textwidth]{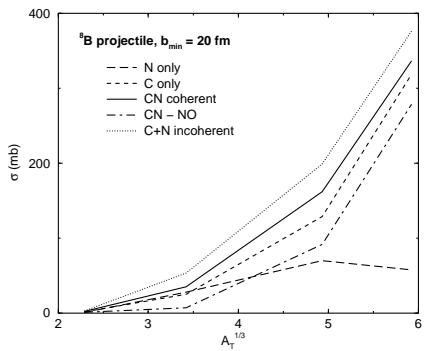}
\caption{\label{Fig5}Total breakup, Coulomb only cross-section and nuclear
only breakup cross-section  for $^{8}$B projectile,
for a lower radial cutoff $b_{\rm min}=20$ fm.}
\end{figure}

\begin{figure}
\includegraphics[width=0.31\textwidth]{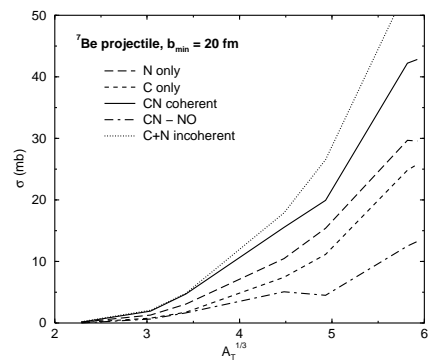}
\caption{\label{Fig6}Total breakup, Coulomb only
and nuclear only  breakup cross-section for $^{7}$Be projectile }
\end{figure}


The Coulomb and nuclear potentials combine together coherently to give breakup,
and the destructive interference between these potentials in the surface region
is well known. If, however, the nuclear and Coulomb breakup cross sections
contribute largely to different partial waves, then the total breakup cross
sections will add incoherently and Eq.~(\ref{gsi-c+n}) should be accurate.

In order to answer this question definitively at least in our test cases, we
have performed further CDCC breakup calculations with both Coulomb and nuclear
transition potentials as in  \cite{nunes99,tost01}, along with sufficient
radial and partial wave limits to encompass all the resulting breakup cross
sections. For reference we also performed `Coulomb only' calculations, where
there are no nuclear potentials at all. Any pure Coulomb calculation needs at
least to be limited by a minimum impact parameter $b_{\rm min}$ to simulate the
effects of nuclear absorption during grazing and closer collisions.

The combined Coulomb and breakup calculation will therefore give a total
\begin{equation}
\sigma_{CN}=\sigma_C+\sigma_N+\sigma_I
\end{equation}
which defines an interference term $\sigma_I$ by the difference with the sum of
Coulomb-only and nuclear-only calculations.
We have found that $\sigma_I$ is sometimes negative (destructive interference),
sometimes positive (constructive interference), and often large. Thus, although one
can construct $\sigma_N$ from some scaling model, the mere subtraction of it from
the data would give a ``contaminated'' Coulomb breakup cross-section:
\begin{equation}
\hat{\sigma}_C=\sigma_C+\sigma_I = \sigma_{CN} - \sigma_{N} \ .
\end{equation}

The use of, say, the equivalent photon method as done in \cite{naka99,datta03}
to extract the $B(E1)$ distribution for $^8$B and $^{11}$Be or for that matter
any other halo nucleus could be questionable if $\sigma_I$ is large. To
ascertain typical sizes of $\sigma_I$, we first show in figures \ref{Fig4},
\ref{Fig5} and  \ref{Fig6} the cross-sections, $\sigma_C$,  $\sigma_N$ and
$\sigma_{CN}$ as the short dashed, long dashed and solid lines respectively,
for the $^{11}$Be and $^8$B at $E_{lab}=44$ MeV/n, and $^7$Be at $E_{lab}=100$ MeV/n,
using $b_{\rm min}=20$ fm in each case.
\begin{figure}
\includegraphics[width=0.33 \textwidth]{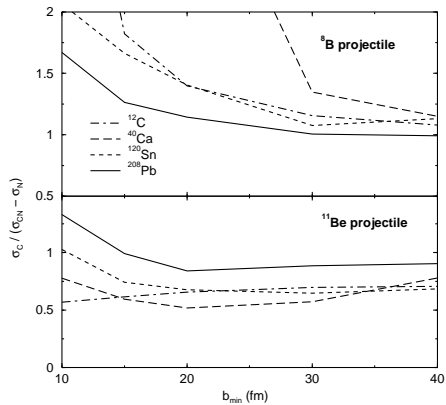}
\caption{\label{Fig7} Ratios of the true to the contaminated Coulomb breakup
cross sections $\sigma_C/\hat{\sigma}_C=   \sigma_C/(\sigma_{CN} - \sigma_{N})$
as a function of the lower radial cutoff $b_{\rm min}$, for four different
targets. Results for $^8$B are shown in the upper panel, and for $^{11}$Be in
the lower panel.}
\end{figure}

To see this, in Figs.~\ref{Fig4} and \ref{Fig5}, we plot as the dot-dashed line
the values of $\sigma_{CN}-\sigma_{N}$, which should `ideally' agree with
$\sigma_C$ as the short dashed line. We also plot as the dotted line the
incoherent sum $\sigma_C+\sigma_N$, whose difference from the $\sigma_{CN}$
solid line indicates the effects of interference. It is evident that, for large
$A_T$, the interference term is destructive for $^8$B (the solid line
$\sigma_{CN} < \sigma_{C+N}$ dotted line), so its neglect may lead to
unrealistically smaller $B(E1)$ distributions. Conversely, it is constructive
for  $^{11}$Be, yielding an unrealistically larger $B(E1)$.

Experimentalists (e.g.~\cite{naka04}) often try to minimise these interference
problems by restrictions to small excitation energies, and/or to large impact
parameters. We focus on the impact parameter restrictions, implemented by means
of a maximum angle $\theta_{\rm max}$ for integrating the cross sections over
the centre of mass angle of the projectile fragments. By semiclassical theory,
this angle is related to $b_{\rm min}$ by  $\theta_{\rm max} =
2\eta/(k b_{\rm min})$. We therefore use the same restrictions on the
calculated breakup cross sections. In Fig.~\ref{Fig7} we show the ratios of the
`true' to `contaminated' Coulomb breakup cross sections
$\sigma_C/\hat{\sigma}_C= \sigma_C/(\sigma_{CN} - \sigma_{N})$ as a function of
the lower radial cutoff $b_{\rm min}$.

Ideally  the ratios should be unity, but in fact we see that rarely do these
values even {\em tend} to unity for large $b_{\rm min}$. Only for the $^8$B
projectile on $^{208}$Pb does this occur, and the deviations from unity are
worse in the $^{11}$Be case.  These deviations from unity indicate either that
the long tail of the $^{11}$Be ground state wave function gives rise to small
but significant deviations from the pure Coulomb results even at impact
parameters $\gtrsim 30$ fm, or that diffraction effects are large enough to
break the semiclassical connection between $b_{\rm min}$ and  $\theta_{\rm
max}$. Equivalent plots for the $^7$Be projectile (not shown) give ratios far
from unity,  from diffraction or refraction causing nuclear breakup fragments
to come out at very forward angles,  implying that there is no `safe' angular
region where Coulomb effects dominate \cite{filomena}. In general, it is clear
that theoretical breakup calculations of Coulomb-nuclear interference are
needed for accurate results for the breakup of $^{11}$Be and $^7$Be on any
target.

In conclusion, we have given evidence through detailed CDCC calculations of the
scaling behaviour of the nuclear breakup cross-section.  This cross-section
does scale as $A^{1/3}_T$, the mass number of the target nucleus, but the
scaling does not follow the geometrical form as in normal nuclei. The nuclear
contribution can be as small as 1/15 of the values calculated in
refs.~\cite{dasso98,dasso99,nag01}. We have further calculated the ``error''
due to the nuclear-Coulomb interference in the extracted $B(E1)$ distribution
if the subtraction $\sigma_{CN}-\sigma_N$ is employed in conjunction with the
virtual photon method. We believe that a full quantum calculation, including
both Coulomb and nuclear potentials on equal footing, such as CDCC, is required
to get credible numbers for the $B(E1)$ distribution of neutron- and
proton-rich (as well as some stable) nuclei.

We acknowledge the support  by the CNPq and FAPESP (Brazilian Agencies), by
NSCL, Michigan State University, the National Science Foundation through grant
PHY-0456656 and the U.K. EPSRC by grant GR/T28577.

\end{document}